\documentclass[english]{article}
\usepackage{courier}
\usepackage[T1]{fontenc}
\usepackage[latin1]{inputenc}
\usepackage{a4}
\usepackage{setspace}
\usepackage[dvips]{graphicx}

\makeatletter

\newcommand{\be}{\begin{equation}}
\newcommand{\ee}{\end{equation}}

\usepackage{graphicx}% Include figure files
\usepackage{dcolumn}% Align table columns on decimal point
\usepackage{bm}% bold math
\usepackage{psfig}

\usepackage{babel}

\usepackage{babel}
\makeatother
\begin{document}

\title{Computational methods in Coupled Electron-Ion Monte Carlo}

\author{Carlo Pierleoni$^{a}$
\footnote{Corresponding author: Carlo Pierleoni, Physics Dept. University of
L'Aquila, Via Vetoio, 67010 Coppito, L'Aquila (Italy), fax:+39-0862433033;
email: carlo.pierleoni@aquila.infn.it} 
and David M. Ceperley$^{b}$}

\maketitle
$^{a}$Department of Physics, University of L'Aquila, Via Vetoio,
I-67010 L'Aquila, Italy

$^{b}$Dept. of Physics and NCSA, University of Illinois at Urbana-Champaign,
Urbana, IL 61801, USA

\begin{abstract}
In the last few years we have been developing a Monte Carlo
simulation method to cope with systems of many electrons and ions
in the Born-Oppenheimer (BO) approximation, the Coupled
Electron-Ion Monte Carlo Method (CEIMC). Electronic properties in
CEIMC are computed by Quantum Monte Carlo (QMC) rather than  by
Density Functional Theory (DFT) based techniques. CEIMC can, in
principle, overcome some of the limitations of the present DFT
based ab initio dynamical methods. Application of the new method
to high pressure metallic hydrogen has recently appeared. In this
paper we present a new sampling algorithm that we have developed
in the framework of the Reptation Quantum Monte Carlo (RQMC)
method chosen to sample the electronic degrees of freedom, thereby
improving its efficiency. Moreover, we show here that, at least
for the case of metallic hydrogen, variational estimates of the
electronic energies lead to an accurate sampling of the proton
degrees of freedom.
\end{abstract}

\newpage

\section{Introduction}

Modern theoretical methods in condensed matter physics and
chemistry rely heavily on numerical simulations. The problem of
solving the Schroedinger equation for many-body systems is too
difficult to be addressed directly, even within the simplification
provided by the Born-Oppenheimer approximation. In the most
popular practical approaches (Hartree-Fock (HF) and the Density
Functional Theory (DFT) based methods\cite{martin04})  the
original problem is replaced by the problem of solving the time
independent Schroedinger equation for a single electron in the
field of the nuclei and the mean field generated by the other
electrons. DFT is, in principle, an exact theory but the energy
functional must be treated approximately for practical purposes.
In the simplest Local Density Approximation (LDA), this exact
theory becomes a self-consistent mean field theory. Extensions of
LDA, such as Generalized Gradient Approximation (GGA) provide more
accurate results but remain essentially at the level of an
effective mean field treatment. Despite the mean field character,
DFT schemes have proved to provide quite accurate results for many
different systems\cite{martin04}

In 1985, Car and Parrinello introduced an efficient method to
couple standard Molecular Dynamics for classical nuclei with the
electronic structure calculation at the level of LDA done {}``on
the fly'' to extract the nuclear forces\cite{cpmd85}. Because the
method allowed study of the statistical mechanics of classical
nuclei with many body electronic interactions, it opened the way
for the use of simulation methods for realistic systems with an
accuracy well beyond the limits of effective force fields
available. In the last twenty years, the number of applications of
the Car-Parrinello ab-initio molecular dynamics has ranged from
simple covalent bonded solids, to high pressure physics, material
science and biological systems. There have also been extensions of
the original algorithm to simulate systems at constant temperature
and constant pressure\cite{bernasconi95}, finite temperature effects for
the electrons \cite{alavikohanoffpf}, and quantum nuclei
\cite{marx-par}.

Despite recent progress, DFT suffers from well-known limitations,
for example, excited state properties such as optical gap and
spectra are less reliable. DFT shows serious deficiencies in
describing van der Waals interactions, non-equilibrium geometries
such as reaction barriers, systems with transition metals and/or
cluster isomers with competing bonding
patterns\cite{martin04,Foulkes01}. As a consequence, current
ab-initio predictions of metallization transition at high
pressures, or even prediction of phase transitions are often only
qualitative. Hydrogen is an extreme
case\cite{maksimivshilov,stadelemartin,johnsonaschroft} but even
in silicon the diamond/$\beta$-tin transition pressure and the
melting temperature are seriously underestimated\cite{alfe04}.

Another route to the ground state properties of a system of many
electrons in presence of nuclei is the Quantum Monte Carlo
method\cite{hammond,Foulkes01}. In its simplest form, an analytic
many electron wave function is chosen on the basis of the
variational principle (Variational Monte Carlo, VMC) and the
quantum averages are obtained by a Metropolis Monte Carlo
simulation of the electronic coordinates. A more accurate
representation of the ground state wave function can be obtained
by projecting the variational wave function with the operator
$exp\{-\beta_{e}H\}$ where $H$ is the many-body hamiltonian, and
$\beta_{e}$ is the projection time. Provided that the variational
wave function is not orthogonal to the ground state wave function,
the projected function tends exponentially fast to the ground
state wave function as $\beta_{e}\rightarrow \infty$. Since matrix
elements of the above projection operator at large values of
$\beta_e$ are unknown for non trivial systems, a Trotter breakup
in many ($P$) small imaginary time intervals
($\tau_{e}=\beta_{e}/P)$ must be employed. In the configuration
representation, each projection corresponds to a $3N$-dimensional
integral which can be performed by Metropolis Monte Carlo method
provided that the propagator in imaginary time can be chosen real
and can be interpreted as a probability distribution. This is the
essence of the Diffusion Monte Carlo method (DMC) which is an
{}``exact'' method for systems of bosons or boltzmannons. This
means that all systematic errors in a simulation are under control
in the sense that they can be reduced as much as desired. Since
electron are fermions, the above scheme fails because the
imaginary time propagator must be completely antisymmetric under
exchange of two electrons and therefore cannot be chosen strictly
non-negative everywhere in configurational space. This is the
origin of the infamous {}``fermion sign problem''. In order to
avoid the sign problem the {}``fixed node approximation'' has been
proposed and used routinely to perform fermion simulations\cite{Foulkes01}. The
energy calculated with this approximation is variational with
respect to the position of the nodal surfaces of the trial wave
function. Over the years, the level of accuracy of the fixed node
approximation for simple homogeneous systems, such as $^{3}He$ and
the electron gas, has been systematically improved by introducing
more sophisticated nodal surfaces (backflow
orbitals)\cite{panoffcarlson,kwon}. In more complex, inhomogeneous
situations such as atoms, molecules and extended systems of
electrons and nuclei, progress have been somewhat slower.
Nonetheless, in most cases, fixed-node QMC methods have proved to
be more accurate than mean field methods (HF and
DFT)\cite{Foulkes01}. Computing ionic forces with QMC to replace
the DFT forces in the ab-initio MD, is more difficult and a
general and efficient algorithm is still missing. 
Moreover, the computer time required for
a QMC estimate of the electronic energy is, in general, more than
for a corresponding DFT-LDA calculation. These problems have
seriously limited the development of an ab-initio simulation
method based on the QMC solution of the electronic problem {}``on
the fly''.

In recent years, we have developed a different strategy based
entirely on the Monte Carlo method both for solving the electronic
problem and for sampling the ionic configuration
space\cite{dewing01,dmc03}.  The new method, called the Coupled
Electron-Ion Monte Carlo method (CEIMC) has been applied so far to
high pressure metallic hydrogen where it has found quite different
effects of temperature than CPMD based on the LDA
forces\cite{pch04}. Our present interpretation of the disagreement
is that LDA provides a Born-Oppenheimer surface quite smoother
than the more accurate QMC one and this strongly affects the
structure of the protonic system at $T > 0$.

The paper is organized as follows. The following section
\ref{sec:The-Coupled-Electron-Ion} is devoted to an outline of the
CEIMC method. We will not go into all details since two long
articles have appeared on general aspects and early
implementations of the method\cite{dewing01,dmc03}. One of the new
aspects that we have recently implemented in CEIMC, not described
in those references, is the Reptation Quantum Monte Carlo
projection of the electronic variational wave function\cite{BaroniMoroni}. 
So in the
subsection \ref{sub:Reptation-Quantum-Monte} we review the RQMC
method and in the following subsection
\ref{sub:The-bounce-algorithm} we focus on the sampling algorithm,
we introduce our new scheme to improve efficiency and reliability
of RQMC and we provide an analytical proof. In
Section\ref{sec:Results} we report numerical results on the
convergence of the new scheme with the projection time and with
the Trotter time step. Finally, in section \ref{sec:Conclusions},
we conclude.

\section{\label{sec:The-Coupled-Electron-Ion}The Coupled Electron-Ion Monte
Carlo method}

CEIMC method is based on the Born-Oppenheimer (BO) separation
between the slow nuclei and the fast electrons. This is in
contrast with other Quantum Monte Carlo methods, Diffusion Monte
Carlo (DMC)\cite{hammond,Foulkes01} or finite temperature Path Integral Monte Carlo
(PIMC)\cite{rmp95,como95} methods where electrons and ions are treated on the same
footing. As usual, the BO approximation allows to overcome the
limitations of the other QMC methods, while introducing an often
negligible error.

In CEIMC, the configurational space of the proton degrees of
freedom at inverse temperature $\beta=(k_{B}T)^{-1}$ is sampled
with a Metropolis algorithm in which the difference between the BO
energy of a proton state $S$ and of a trial state $S'$ is computed
by an electronic ground state QMC calculation. The QMC estimate of
the energy difference $\Delta=[E(S')-E(S)]$ has statistical noise
which would bias the standard Metropolis algorithm. Unbiased
sampling of the proton configurations is achieved by the penalty
method\cite{penalty} which replaces the energy difference $\Delta$
in the acceptance formula by $\Delta+(\beta\sigma_{\Delta})^{2}/2$, where
$\sigma_{\Delta}^{2}$ is the variance of the energy difference. Since
$\sigma_{\Delta}^{2}>0$, the noise always causes extra rejections but this
compensates for {}''uphill'' moves accepted because of a favorable
energy fluctuation.

Several methods for computing energy differences in QMC are
available.\cite{dewing01,dmc03}. A simple and efficient method is
to sample the electronic degrees of freedom from a distribution
function which is the sum of the electronic distribution functions
for the $S$ and $S'$ states (\textit{e. g.} the sum of the the
square of the trial wave functions in VMC). Averages of operators
involving electronic degrees of freedom and a single proton
configuration, say $S$, (for instance total energy, variance,
etc.) are then computed by correlated
sampling\cite{hammond,dewing01,dmc03}. For the typical size of the proton
moves (between $0.01\textrm{Å}$ and $0.5\textrm{Å}$ for classical
protons depending on density and temperature) and the typical
system size (up to 54 protons) we have investigated, this method
is much more efficient than performing two independent electronic
calculations for the state $S$ and $S'$.

In the ground state QMC methods, an electronic trial wave function
must be chosen according to the physics of the system being
studied. For the metallic phase of hydrogen, we have recently
developed analytic functions which include backflow and three-body
correlations\cite{hcpe03}. These wave functions are particularly
appropriate to CEIMC since they have accurate energies already at
the variational level, they have no adjustable parameters
requiring optimization, and their computational cost is much less
than using orbitals expanded in a plane wave basis typically used
in QMC calculations\cite{natoli93}.

In metallic systems, finite size effects are large and must be
suitably treated. The common procedure is to repeat the
calculation for systems of increasing size and extrapolate to the
thermodynamic limit but this is impractical within CEIMC, since it
would have to be performed for any proposed protonic step before
its acceptance. A much better strategy is to use Twist Averaged
Boundary Condition (TABC) \cite{lzc01,dmc03} which reduces the
finite size error in the energy to the classical $1/N$ behavior.
It consists in averaging the energy over the phase that the many
body wave function can pick if a single electron wraps around the
super-cell. This is equivalent to Brillouin zone sampling in the
single electron approximation.  Within CEIMC it does not cause a
large increase in required CPU time/step.

Finally a recent improvement of the method is the introduction of
quantum effects for the protons, quite important in high pressure
hydrogen. This is done by developing the thermal density matrix of
protonic degrees of freedom on the BO surface in Feynman Path
Integrals\cite{rmp95,pch04}. A similar technique in the context of
Car-Parrinello method has appeared\cite{marx-par}. We are not
going to discuss the last two aspects of the CEIMC. While TABC
implementation in CEIMC has been described in ref.\cite{dmc03},
our implementation of PIMC for proton degrees of freedom in CEIMC
will be the subject of a future publication.

\subsection{\label{sub:Reptation-Quantum-Monte}Reptation Quantum Monte Carlo
Method}

To go beyond VMC electronic energies, we implemented a Reptation
Quantum Monte Carlo algorithm (RQMC)\cite{BaroniMoroni}, rather
than Diffusion Monte Carlo algorithm (DMC). The implementation of
the energy difference method is more straightforward in RQMC, nor
are averages of observables which do not commute with the
hamiltonian biased.

In RQMC the ground state wave function is obtained by constructing
an imaginary time path integral for the electronic degrees of
freedom. If $|\Psi_{0}\rangle$ is the trial state, the trial state
projected in a {}``time'' $\beta_{e}/2$,
$|\Psi_{\beta_{e}/2}\rangle=e^{-\beta_{e}H/2}|\Psi_{0}\rangle$,
will converge to the ground state for large $\beta_{e}$. Let us
define the {}``partition'' function 
\begin{equation}
Z_{\beta_{e}}=\langle\Psi_{0}|e^{-\beta_{e}H}|\Psi_{0}\rangle.\label{eq:zetabeta}\end{equation}
The energy is then defined as \begin{equation}
E(\beta_{e})=-\frac{d}{d\beta_{e}}\ln
Z_{\beta_{e}}=\frac{1}{Z_{\beta_{e}}}\langle\Psi_{0}|e^{-\beta_{e}H}H|\Psi_{0}
\rangle=\langle E_{L}(R)\rangle\label{eq:energy}
\end{equation}
where the averages of the local energy,
$E_{L}(R)=\Re(\Psi_{0}^{-1}(R)\, H\Psi_{0}(R))$, are with respect
to the path average. In practice, the energy is computed as the
average of the local energy at the two ends of the path. Here
$\Re$ indicates the real part in the case that the trial function
or Hamiltonian is complex. The energy $E(\beta_{e})$ is an upper
bound to the fixed node energy for each value of $\beta_e$, it
converges to this at large $\beta_{e}$, and its $\beta_{e}$
derivative is strictly negative. This latter quantity is, in fact,
minus the variance of the total energy
\begin{equation}
\sigma^{2}(\beta_{e})=-\frac{dE(\beta_{e})}{d\beta_{e}}=\langle
E_{L}(0)E_{L}(\beta_{e})\rangle -\langle E(\beta_{e})\rangle^{2}
.\label{eq:variance}
\end{equation} 
The variance tends to zero for
large enough values of $\beta_{e}$ providing a useful signal for
the convergence of the energy to the ground state. This is the
zero variance theorem in RQMC. 
%If we assume an exponential dependence of the energy 
%$E(\beta)=E_0+b e^{(-\beta_e\delta)}$, where $E_0, b$ and $\delta$ are numerical
%parameters,using eq.(\ref{eq:variance}) we obtain
%\begin{equation}
%E(\beta_{e})= E_0+\frac{\sigma^2(\beta_e)}{\delta}
%\end{equation}
%that is a linear dependence of the energy versus the variance which can be used to
%reliably extrapolate for $E_0$. Also the inverse slope is related to the energy gap of
%the states being projected in the imaginary time dynamics.
On the other hand, the variance of
the local energy computed at either end of the path is the mixed
estimator of DMC for $\sigma^{2}(\beta_{e})$. In practical
implementations, it is desirable to keep $\beta_{e}$ as small as
possible to maximize the efficiency of the energy difference
method.

To compute the needed density matrix elements, we divide the
projection time $\beta_{e}$ into $P$ time slices
$\tau_{e}=\beta_{e}/P$ and make a semi-classical approximation for
$\exp(-\tau_{e}H)$. Our notation for a single electronic
configuration is $R=\left\{
\mathbf{r}_{1},\ldots,\mathbf{r}_{N}\right\} $, while for the
entire path is $s=\left\{ R_{0},R_{1},\ldots,R_{P}\right\} $. The
probability distribution for a path is \begin{equation}
\Pi(s)=\exp\left\{
-U(R_{0})-U(R_{p})-\sum_{i=1}^{P-1}L_{s}(R_{i+1},R_{i})\right\}
\label{eq:pdf}\end{equation} where $U(R)=\Re[ln\Psi{}_{0}(R)]$ and
$L_{s}(R,R')$ is the symmetrized link action for our approximation
of the short time propagator. We have used the importance sampling
Green's function of the DMC propagator 
\begin{equation}
\left\langle R|e^{-\tau_{e}H}|R'\right\rangle
=\left|\frac{\Psi_{0}(R)}{\Psi_{0}(R')}\right|\,\exp\left[-\tau_{e}E_{L}(R)-\frac{[R'-R-2\lambda\tau_{e}F(R)]^{2}}{4\lambda\tau_{e}}\right].
\label{eq:action}\end{equation} 
where the force is
$F(R)=\mathbf{\nabla}U(R)$ and $\lambda=\hbar^{2}/2m_{e}$, which
provides the symmetrized link action 
\begin{eqnarray}
L_{s}(R,R') & = & \frac{\tau_{e}}{2}\left[E_{L}(R)+E_{L}(R')
+\lambda\left(F^{2}(R)+F^{2}(R')\right)\right] \nonumber \\
& + & \frac{(R-R')^{2}}{4\lambda\tau_{e}}+\frac{(R-R')\cdot(F(R)-F(R'))}{2}
\label{eq:simaction}
\end{eqnarray}
An alternative form for the link action could be obtained through
the pair action developed in finite temperature Path Integral
MC\cite{rmp95}. However, we have not implemented this form and do
not have a comparison of its efficiency.

In order to impose the fixed phase constraint on the projected
wave function, we must add to the link action a term of the form
$L_{s}^{FP}(R,R')=\lambda\tau_{e}\int_{0}^{1}d\eta\left|\nabla\phi(X(\eta))\right|^{2}$
where $\phi(X)$ is the phase of the trial wave function at
electronic position $X$ and the integral is taken over all paths
$X(\eta)$ with boundary conditions $X(0)=R,\, X(1)=R'$. We have
taken an end-point approximation for this term except for real
wave functions in which case fixed-node boundary conditions were
used.

Note that in the expressions above, the dependence on the nuclear
degrees of freedom was not shown even though all quantities depend
on them. The probability distribution of an electronic path will
be $\Pi(s,S)$ where $S$ indicates the position of all nuclei.
Because we have an explicit distribution of the electronic paths,
it is straightforward to apply the importance sampling scheme for
the energy differences by sampling the probability distribution
$[\Pi(s,S)+\Pi(s,S')]$ where $S$ and $S'$ are the current and the
trial protonic state, respectively. Note also, for VMC $\Pi(s,S)
\propto \left|\Psi_{0}(s,S)\right|^{2}$ becomes the square of
modulus of the trial wave function (no projection and
$R_{0}=R_{P}$).

\subsection{\label{sub:The-bounce-algorithm}The {}``bounce'' algorithm}

In the original work on RQMC\cite{BaroniMoroni}, the electronic
path space was sampled by a reptation algorithm, an algorithm
introduced to sample the configurational space of linear polymer
chains. The slithering snake or reptation method seems to have
originated by Kron\cite{kron} and by Wall and
Mandel\cite{wallmandel}. Given a path configuration $s$, a move is
done in two stages. First one of the two ends (either $R_{0}$ or
$R_{P}$) is sampled with probability 1/2 to be the growth end
$R_{g}$. Then a new point near the growth end is sampled from a
Gaussian distribution with center at
$R_{g}+2\lambda\tau_{e}F(R_{g})$. In order to keep the number of
links on the path length constant, the old tail position is
discarded in the trial move. The move is accepted or rejected with
the Metropolis formula based on the probability of a reverse move.
For use in the following, let us define the direction variable $d$
as $d=+1$ for a head move ($R_{g}=R_{P}$), and $d=-1$ for a tail
move ($R_{g}=R_{0}$). In standard reptation, the direction $d$ is
chosen randomly at each attempted step.

In the standard reptation algorithm,  the transition probability
$P(s\rightarrow s')$ is the product of an attempt probability
$T_{d}(s\rightarrow s')$ and an acceptance probability
$a_d(s\rightarrow s')$. Note that the path distribution given in
Eq.(\ref{eq:pdf}), because of the symmetrized link action does not
depend on the direction $d$ in which it was constructed. In the
Metropolis algorithm, the acceptance probability for the attempted
move is
\begin{equation}
a_d(s\rightarrow s')=min\left[1,\frac{\Pi(s')T_{-d}(s'\rightarrow
s)}{\Pi(s)T_{d}(s\rightarrow
s')}\right]\label{eq:acceptance}\end{equation} which ensures that
the transition probability $P_d(s \rightarrow s')$ satisfies
detailed balance\begin{equation} \Pi(s)P_d(s\rightarrow
s')=\Pi(s')P_{-d}(s'\rightarrow s)\label{eq:DB}\end{equation}

The autocorrelation time of this algorithm in Monte Carlo steps,
that is the number of MC steps between two uncorrelated
configurations, scales as $[(\beta_{e}/\tau_{e})^{2}/A]$, where
$A$ is the acceptance rate, an unfavorable scaling for large
$\beta_{e}$. Moreover the occasional appearance of persistent
configurations bouncing back and forth without really sampling the
configuration space has been previously observed\cite{saverio}.
These are two very unfavorable features, particularly in the
present context, where we need to perform many different
electronic calculations (at least one per protonic move). There is
a premium for a reliable, efficient and robust algorithm.

We have found that a minimal modification of the reptation
algorithm solves both of these problems. The idea is to chose
randomly the growth direction at the beginning of the Markov
chain, and reverse the direction upon rejection only, the
{}``bounce'' algorithm. As far as we are aware, ''bounce''
dynamics has not been previously investigated for RQMC, though
Wall and Mandel\cite{wallmandel} mentioned it without a detailed
proof and subsequent polymer simulations did not use bounce,
perhaps because the acceptance ratio in the polymer systems is
much smaller than in RQMC. There is a related algorithm for
directed loop algorithm on the lattice and for simulations of
trapped diffusion\cite{anlauf,gallos}.
%\footnote{Comment from Mattias Troyer: This method is not well known but seems
%related to the \char`\"{}directed loop\char`\"{}algorithm of Sandvik
%and Syljuasen (cond-mat/0211513 andcond-mat/0202316). They try to
%move in one direction as long as possible and reverse the direction
%(bounce) only when necessary. It would be very interesting to study
%the relationship between your method applied to worms and the directed
%loop algorithm. However those algorithms work in the space of individual
%worldlines, and explicitly satisfy detailed balance. }.

What follows is the proof that the bounce algorithm samples the
correct probability distribution $\Pi(s)$. The variable $d$ is no
longer randomly sampled, but, as before, the appropriate move is
sampled from the same Gaussian distribution $T_{d}(s\rightarrow
s')$ and accepted according to the Eq. (\ref{eq:acceptance}). To
be able to use the techniques of Markov chains, we need to enlarge
the state space with the direction variable $d$. In the enlarged
configuration space $\left\{ s,d\right\} $, let us define the
transition probability $P(s,d\rightarrow s',d')$ of the Markov
chain.  The algorithm is a Markov process in the extended path
space,  and it is ergodic as DMC method, hence, it must converge
to a unique stationary state, $\Upsilon(s,d)$ satisfying the
eigenvalue equation:
\begin{equation}
\sum_{s,d}\Upsilon(s,d)\, P(s,d\rightarrow
s',d')=\Upsilon(s',d')\label{eq:eigenvalue}.\end{equation} We show
that our desired probability $\Pi(s)$ is solution of this
equation. Within the imposed rule not all transitions are allowed,
but $P(s,d\rightarrow s',d')\neq 0$ for $d=d'$ and $s\neq s'$
(accepted move), or $d'=-d$ and $s=s'$ (rejected move) only.
Without loss of generality let us assume $d'=+1$ since we have
symmetry between $\pm1$. Eq. (\ref{eq:eigenvalue}) with
$\Upsilon(s,d)$ replaced by $\Pi(s)$ is\[
\Pi(s')P(s',-1\rightarrow s',1)+\sum_{s\neq
s'}\Pi(s)P(s,1\rightarrow s',1)=\Pi(s').\] Because of detailed
balance Eq.(\ref{eq:DB}), we have $\Pi(s)P(s,1\rightarrow
s',1)=\Pi(s')P(s',-1\rightarrow s,-1)$, which when substituted in
this equation gives\[ \Pi(s')\left[P(s',-1\rightarrow
s',1)+\sum_{s}P(s',-1\rightarrow s,-1)\right]=\Pi(s').\] Note that
we have completed the sum over $s$ with the term $s=s'$ because
its probability vanishes. The term in the bracket exhausts all
possibilities for a move from the state $(s',-1)$, thus it adds to
one. Hence $\Pi(s)$ is a solution of eq. (\ref{eq:eigenvalue}) and
by the theory of Markov chains, it is the probability distribution
of the stationary state.

\section{\label{sec:Results}Results}

\begin{figure}
\includegraphics[%
  clip,
  scale=0.4]{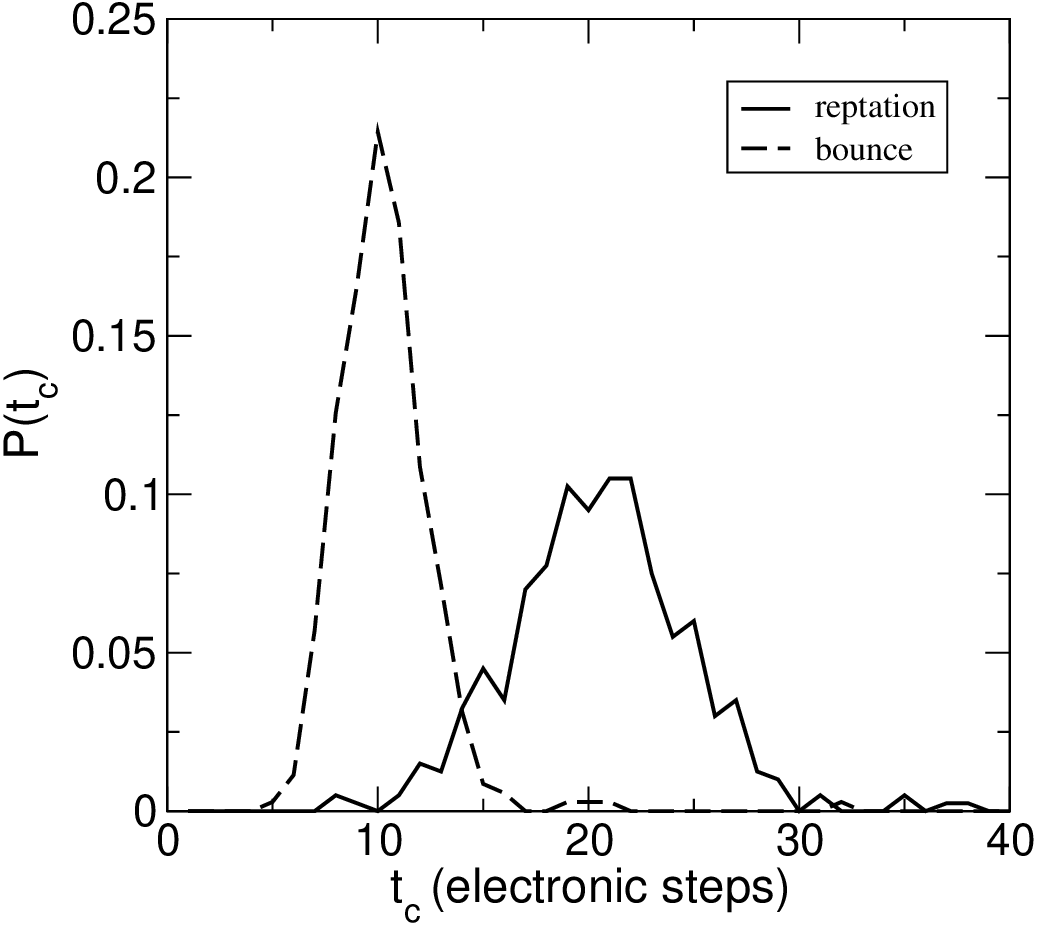}

\caption{\label{cap:fig1}Histogram of the correlation time $t_{c}$
of the energy difference. Comparison between the reptation and the
bounce algorithm for a path with 4 links.}
\end{figure}

In order to check the validity of our proof we first applied the
bounce algorithm to an analytically solvable model, namely a one
dimensional harmonic oscillator and obtained the expected results.
For a realistic test, we compare the standard and the bounce
algorithms for a fixed pair of protonic configuration $(S,S')$
generated during a VMC run of liquid hydrogen at $r_{s}=1.31$ and
$T=5000K$. We have considered $N_{p}=N_{e}=16$ protons and
electrons using analytic wave functions with 3-body and backflow
terms at the $\Gamma$ point (periodic boundary
conditions)\cite{hcpe03}. In the test, we fixed the electronic
imaginary time step to $\tau_{e}=0.04\, h^{-1}$ and the projection
time to $\beta_{e}=0.2\, h^{-1}$ which corresponds to 4 links. The
key quantity in CEIMC is the correlation time $t_{c}$ in
electronic MC steps of the energy difference
$\Delta=E_{BO}(S')-E_{BO}(S)$ which determines, for a fixed length
of the electronic run and for a given proton displacement, the
noise level. The shorter the correlation time $t_{c}$ the larger
the number of independent determinations of the energy difference.
This implies smaller noise level and a larger acceptance for
protonic moves, i.e. a higher efficiency of the algorithm. In fig.
\ref{cap:fig1} we compare the histogram of the correlation time
$t_{c}$ of the energy difference obtained with standard reptation
and with the bounce algorithm over 400 blocks of $10^{5}$
electronic steps. In both calculations the electronic acceptance
rate is 0.89 but the noise level is $0.28$ with standard reptation
and only $0.14$ with the bounce algorithm, in agreement with the
observed correlation times. Note that, not only the average, but
also the width of the distribution is roughly twice as large with
standard reptation than with bounce dynamics.

\begin{figure}
\includegraphics[%
  clip,
  scale=0.45]{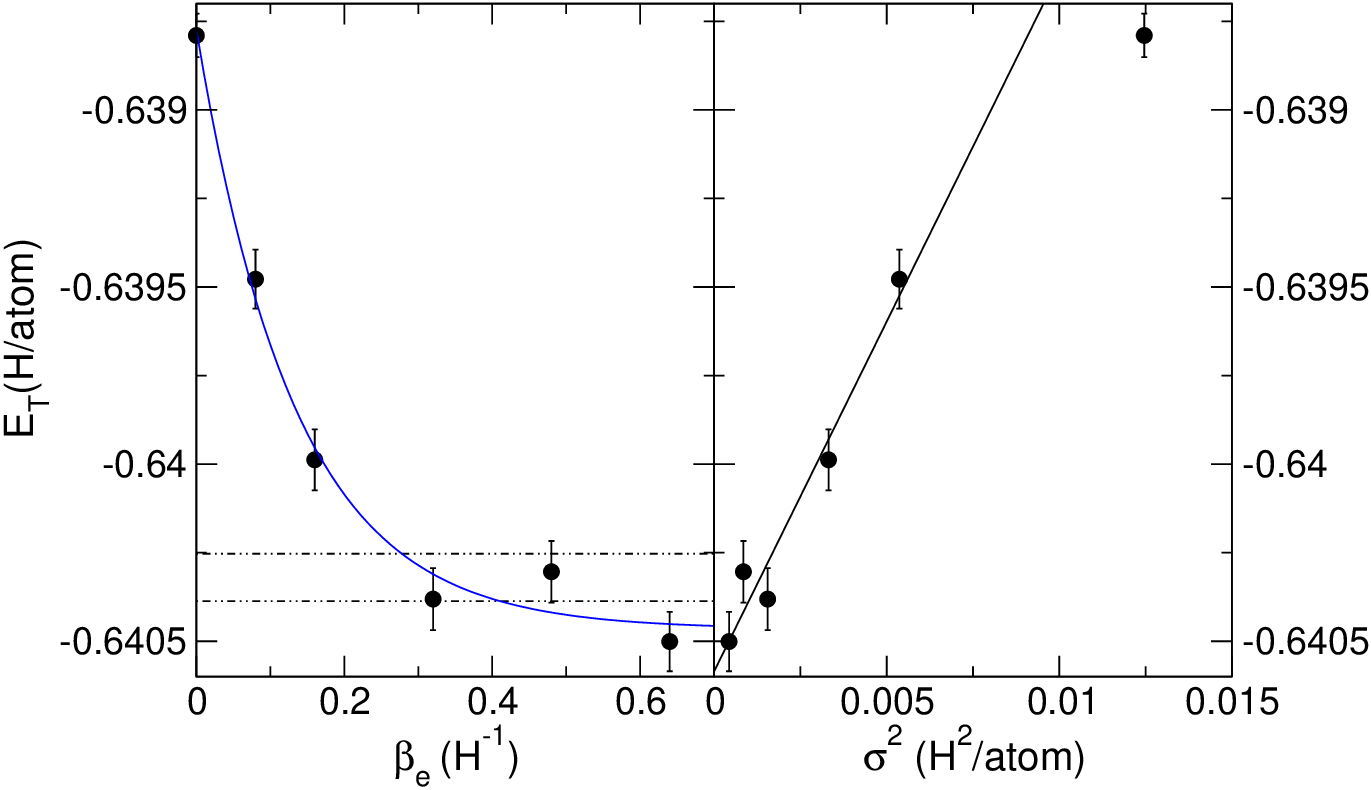}
\caption{\label{cap:nconv}Bcc hydrogen at $r_{s}=1.31$. Convergence with $\beta_e$ 
for the energy (left panel) and total energy versus the variance 
$\sigma^2$ (right panel). In the left panel the curve is a shifted exponential fit while
the horizontal dot-dashed lines represent the DMC result with its statistical error.
%In the right panel the slope of the linear fit for $\sigma^2\le0.005$ 
%($0.20$ in the present case) is related to the
%inverse of the energy gap to the first excited state (see eq.(\ref{eq:variance}))
.}
\end{figure}

Next we study the convergence of the bounce algorithm with respect
to $\tau_{e}$ and $\beta_{e}$. We first consider protons on a bcc
lattice to study the convergence of total energy and variance and
to compare with DMC. As above, we consider $N_e=N_p=16$ at
$r_{s}=1.31$, with the boundary condition
$\theta=2\pi(0.4,\,0.5,\,0.6)$. Data obtained with runs of
$10^{6}$ electronic steps, are shown in fig. \ref{cap:nconv}. At fixed
$\beta=0.16 H^{-1}$ we observed a roughly linear convergence (from below) 
of the total energy with $\tau_e$ (not shown). The results in fig.
\ref{cap:nconv} are for $\tau_{e}=0.04H^{-1}$ which may
underestimate the energy by $0.3mH/atom$. Because of the high
quality of the trial function, the ground state is reached with a
very small projection time. Already at $\beta_{e}=0.6$ the energy
saturates at the value obtained with DMC (essentially infinite
projection time, it is shown as a horizontal line in the upper
left panel). The remarkable linear dependence of the energy versus
the variance below $\sigma^{2}=0.005$ (upper middle panel) can be
used to reliably extrapolate the energy to the $\beta\rightarrow
\infty$ limit.

\begin{figure}
\includegraphics[%
  scale=0.5]{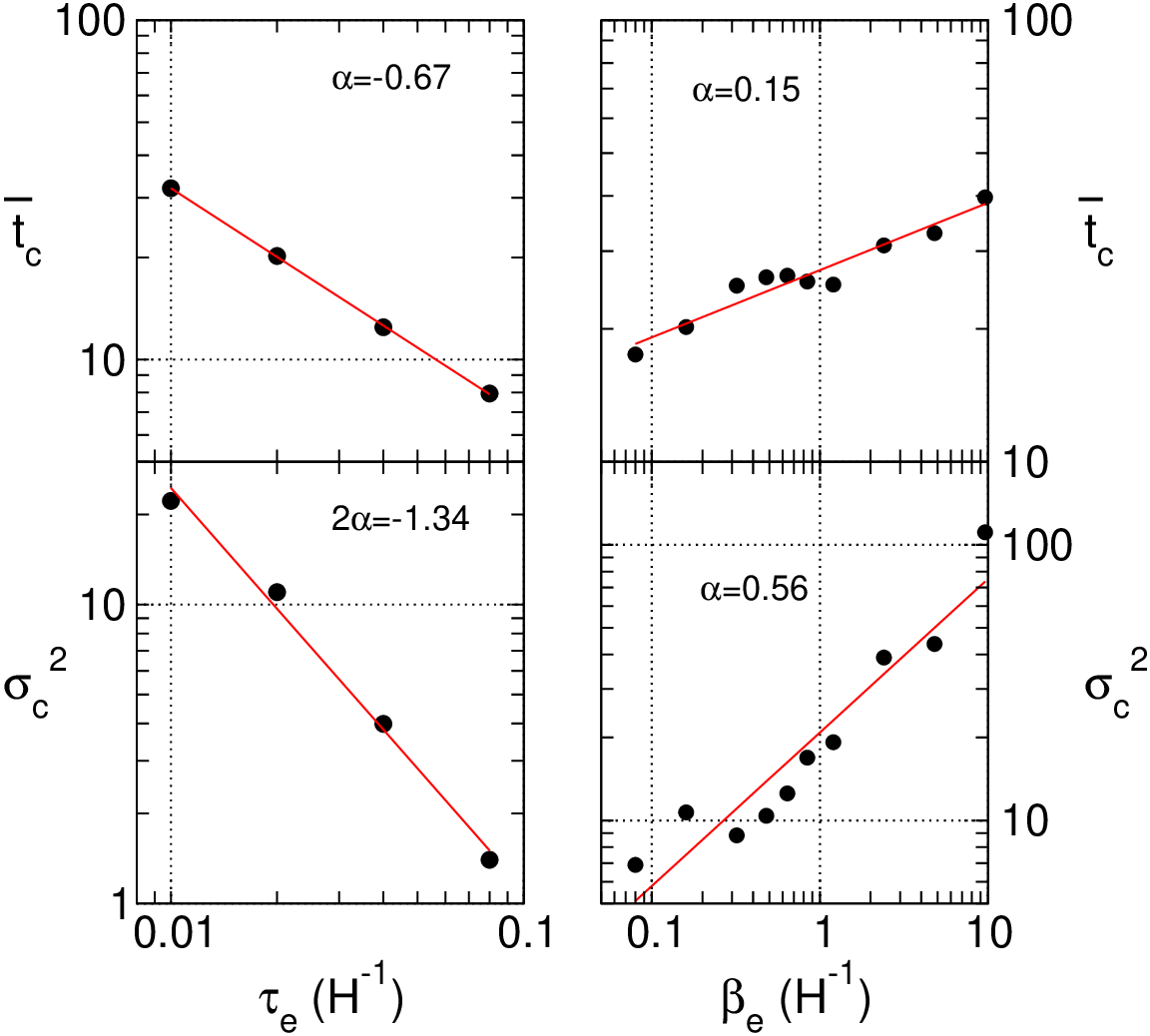}
\caption{\label{fig:ctime_fit}Scaling of the average correlation
time $\overline{t_{c}}$ of the energy difference and of its
variance $\sigma_{c}^{2}$ for a fixed pair of proton
configurations. Left panels show the behavior with $\tau_{e}$ at
fixed $\beta_{e}=0.16H^{-1}$, while right panels show the
$\beta_{e}$ dependence at fixed $\tau_{e}=0.02H^{-1}$. }
\end{figure}

In order to study the convergence of the energy difference and to
estimate the scaling of $t_{c}$ with $\beta_{e}$, we consider a
pair of successive protonic configurations for the same system
generated during a CEIMC run  at T=5000K, i. e. in the liquid
state. At fixed $\beta_{e}=0.16H^{-1}$, we study the convergence
with $\tau_{e}$ in the range
$0.01H^{-1}\leq\tau_{e}\leq0.08H^{-1}$, and at fixed
$\tau_{e}=0.02H^{-1}$, we study the convergence with $\beta_{e}$
in the range $0.08H^{-1}\leq\beta\leq 9.6H^{-1}$ which corresponds
to $4 \leq P\leq 480$ time slices. In figure \ref{fig:ctime_fit}
we show the first two moments $\overline{t_{c}}$ and
$\sigma_{c}^{2}$ of high quality Gaussian fits to the histograms
of $t_{c}$. At fixed $\beta_{e}$, the rejection rate increases
linearly with $\tau_{e}$ (not shown). However, successful moves
are more effective and this results in the observed scaling
$\overline{t_{c}}\sim\sigma_{c}\sim\tau_{e}^{-0.67}$ (left panels)
at least in the limited range of values of $\tau_{e}$ spanned. The
behavior for increasing $\beta_{e}$ at fixed $\tau_{e}$ is more
sluggish. Note that the rejection rate, $0.037$ in the present
case, does not depend on $\beta_{e}$. Both $\overline{t_{c}}$ and
$\sigma_{c}^{2}$ exhibit a somewhat erratic behavior but the
overall scalings are quite favorable. Note that $\sigma_{c}$
appears to scale roughly as $\overline{t_{c}}^{2}$. Although the
quality of the Gaussian fit remains good even at large
$\beta_{e}$, for $\beta_{e}>1$ the histogram of $t_{c}$ starts
developing an small asymmetry with respect to the maximum with
slower decay at large values of $t_{c}$.

%DROP THIS? Data can be represented with a shifted Poisson
%distribution,
%$P(t_{c})=(t_{c}-a_{0})^{n}e^{-(t_{c}-a_{0})/a_{1}}/n!\,
%a_{1}^{n+1}$ for $t_{c}\geq a_{o}$ and $P(t_{c})=0$ for
%$t_{c}<a_{o}$, with $n=7$ (independent of $\beta_{e}$) and fitted
%in terms of the parameters $a_{0}$ and $a_{1}$ achieving a quality
%superior or comparable to the gaussian fit. Remembering that the
%skewness, $\kappa_{3}=\mu_{3}/\mu_{2}^{3/2}$, and the excess
%kurtosis, $\kappa_{4}=(\mu_{4}/\mu_{2}^{2}-3)$ , of the Poisson
%distribution, defined in terms of the central moments
%$\mu_{m}=\langle (t_{c}-\overline{t_{c}})^{m}\rangle$, are
%$\kappa_{3}=2/\sqrt{n+1}$ and $\kappa_{4}=6/(n+1)$, for $n=7$ we
%have $\kappa_{3}=1/\sqrt{2}$ and $\kappa_{4}=3/4$ which indicate
%indeed little deviation from normality.

%The favorable scaling with $\beta_{e}$ observed in figure
%\ref{fig:ctime_fit} does not necessarily hold for other electronic
%properties. However, the energy difference is the key quantity in
%CEIMC since it is used to sample the BO surface and the proposed
%{}``bounce'' algorithm appears to be particularly suitable for
%this method.

%
\begin{figure}
\includegraphics[%
  clip,
  scale=0.35]{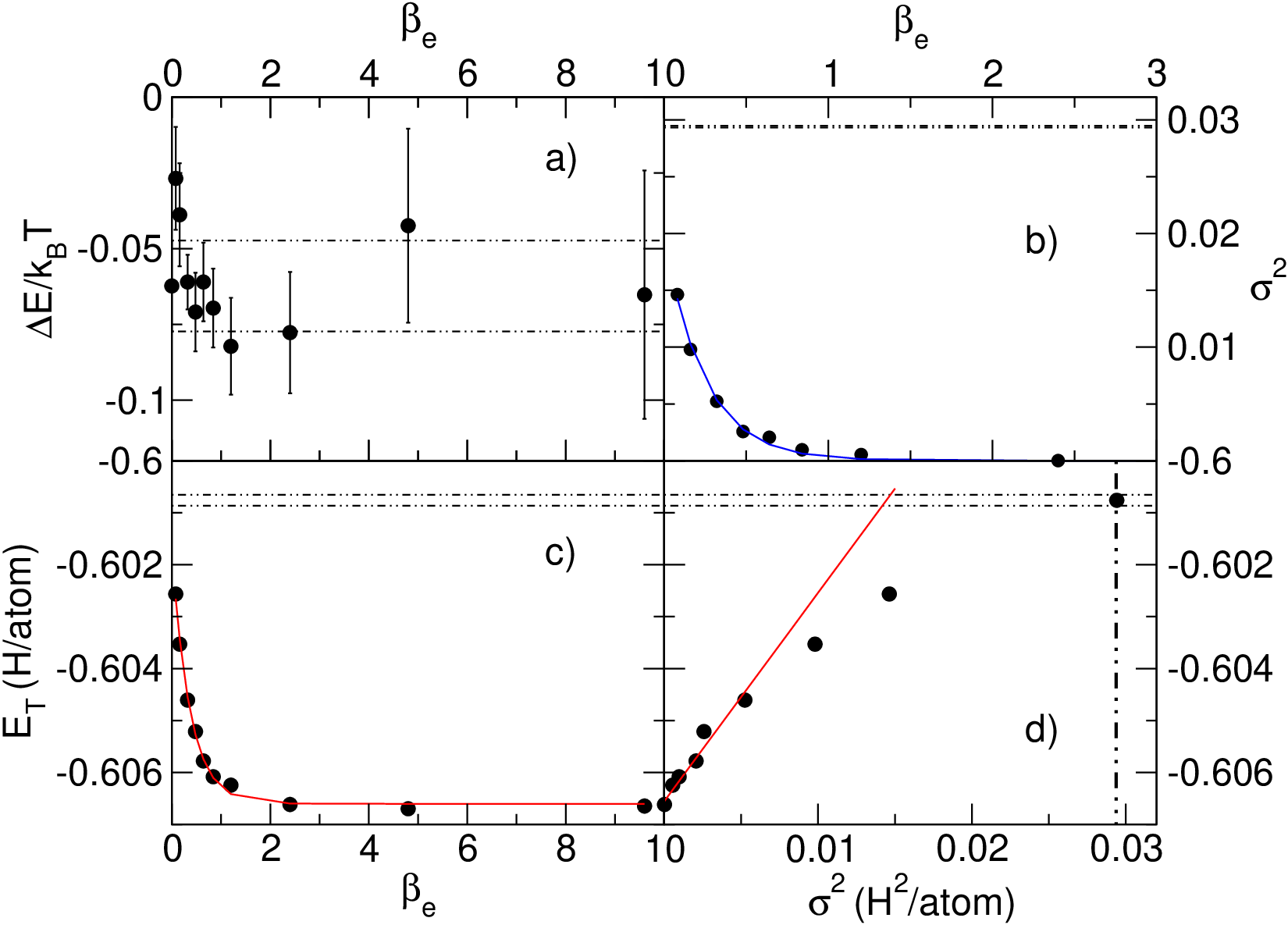}

\caption{\label{fig:tau0.02}$\beta_{e}$ dependence of total energy, variance
and energy difference for a pair of proton configurations $(S,S')$.
The study is performed for $\tau_{e}=0.02H^{-1}$. Dot-dashed lines
represent the variational estimates with their error bars. In panel
b) and c) the lines are exponential fits to data and in panel d) the
continuous line is a linear fit in the region $\sigma^{2}\leq0.005$.
%Extrapolated values for the ground state energy are $E_{0}=-0.60661H/atom$,
%from the exponential fit in panel c), and $E_{0}=-0.60655H/atom$
%from the linear fit in panel d).
}
\end{figure}

In fig. \ref{fig:tau0.02} we report the related energy convergence
study at fixed $\tau_{e}=0.02H^{-1}$. In all panels, horizontal
lines represent the variational estimate with its statistical
error. In particular the panel a) shows that the energy difference
$\Delta E/k_{B}T$ used in CEIMC to perform the
acceptance/rejection test is roughly independent of $\beta_{e}$
(neither is there any $\tau_{e}$ dependence at fixed
$\beta_{e}$)). This result suggests that difference of the
electronic energies at the variational level is accurate enough to
perform CEIMC, at least in the present case of metallic hydrogen
with these analytical trial functions; we can sample the proton
coordinates using VMC and compute the corrections to the energy
and to the equation of state with RQMC for well equilibrated,
statistically independent configurations. From panel c) we see
that the projected energy is lower by $5.7mH/atom=1809K/atom$ with
respect to the variational estimate, a significant change on the
proton energy scale. In panels b) and c) are shown exponential
fits to the data. %We note that the observed decay rate for
%$\sigma$ is roughly three times smaller than the decay rate of the
%energy (do we have an explanation for this?).
Panel d) shows the
energy vs the variance. As previously noticed, linear behavior is
obtained for $\sigma^{2}\leq 0.005$.

Finally, in order to test whether the VMC and RQMC computed BO
surfaces have the same shape in the relevant part of the proton
configurational space and not only at a single point, we have
studied a system of $N_{p}=N_e=54$ atoms at $r_{s}=1$ and
$T=1000K$ with zero twist phase ($\Gamma$ point). Comparison of
our VMC pair correlation functions with CPMD-LDA
results\cite{kh95} at this thermodynamic point has recently
appeared \cite{pch04}. The RQMC calculation has been performed
with $\tau_{e}=0.02H^{-1}$ and $\beta_{e}=1.0H^{-1}$ and provides
an energy of $-0.41114(8)H/atom$ to be compared with the
variational estimate of $-0.4087(1)H/atom$. The computed pressure
is $17.47(1)$Mbars and VMC and RQMC estimates are in agreement
within error bars. Average correlation time and variance of the
energy difference are $\overline{t_{c}}=7.1,\,\sigma_{c}^{2}=2.3$
and $\overline{t_{c}}=16.5,\,\sigma_{c}^{2}=26.5$ for VMC and RQMC
respectively. Therefore, going from VMC to RQMC with the same
efficiency requires electronic runs between two and three times
longer.

%DROP ? In fig. \ref{fig:rs1t1n108gr} we compare the pair
%correlation functions from VMC and RQMC. They are in very good
%agreement except for the $g_{ep}(r)$ and $g_{ee}(r)$ spin-like at
%very short distance (below $0.2a$). This is a finite time step
%error of our approximation of the short time propagator, could be
%improved, and although unpleasant to the view, it is probably
%irrelevant to physical quantities.

%
%\begin{figure}
%\includegraphics[%  clip,  scale=0.5]{rs1t1n108gr.eps}
%\caption{\label{fig:rs1t1n108gr}Pair correlation functions at $r_{s}=1,$
%T=1000K for a system of $N_{p}=54$ protons and $N_{e}=54$ spin-unpolarized
%electrons at $\Gamma$ point. Comparison between VMC and RQMC results. }
%\end{figure}

\section{\label{sec:Conclusions}Conclusions}

In conclusion, we have developed a new sampling algorithm for
reptation Quantum Monte Carlo which we have shown to be more
efficient than the standard sampling scheme and to have a
favorable scaling with the projection (imaginary) time. This new
scheme, which requires a minimal change of existing codes, allows
one to sample long electronic paths with a limited effort. We did
not observe the occurrence of pathological situations previously
reported with the standard scheme where the direction was
resampled each move. We have implemented the new sampling
algorithm in the CEIMC method and found that the correlation time
of the energy difference for a given pair of protonic
configurations grows like the projection time to the power 0.15.
This means, in practice, that the noise level in CEIMC will get
only moderately worse with increasing projection time, i.e.
approaching the ground electronic state. More important, we have
found that the difference in energy between the two configurations
is not sensitive to the projection time, suggesting that CEIMC
sampling with VMC provides accurate dynamics. This conjecture has
been verified for metallic hydrogen at a single thermodynamic
point.

An interesting question that remains unanswered is how general our
conclusions are. Since the trial wave functions used in the
present application are particularly accurate, which is not
generally the case, caution must be exercised in applying the
algorithms to cases where the accuracy of the trial function is
unknown.

Early aspects of the CEIMC algorithm were developed in collaboration
with M. Dewing. We have the pleasure to thank S. Moroni for useful
discussions. This work has been supported by a visiting grant from
INFM-SezG and by MIUR-COFIN-2003. Computer time has been provided
by NCSA (Illinois), PSC (Pittsburg) and CINECA (Italy) through the
INFM Parallel Computing initiative.

\end{document}